\documentclass[preprint,aps,amsmath,byrevtex,prd,titlepage,
nofootinbib]{revtex4-2}

\usepackage{dcolumn}
\usepackage{amsmath}
\usepackage{amssymb}
\usepackage{bm}
\usepackage{hyperref}
\usepackage{graphicx}
\usepackage{slashed}
\usepackage{subfigure}

\newcommand{\beq}{\begin{equation}}
\newcommand{\eeq}{\end{equation}}
\newcommand{\eq}[1]{Eq.~(\ref{#1})}

\begin{document}

\title{One more radiative-recoil correction to the Lamb shift in muonium}

\author {Michael I. Eides}
\email[Email address: ]{meides@g.uky.edu}
\affiliation{Department of Physics and Astronomy,
University of Kentucky, Lexington, KY 40506, USA}
\author{Valery A. Shelyuto}
\email[Email address: ]{shelyuto@vniim.ru}
\affiliation{D. I.  Mendeleyev Institute for Metrology,
St.Petersburg 190005, Russia}

\begin{abstract}
We calculate radiative-recoil contribution   of order $Z^2\alpha(Z\alpha)^5(m/M)^2m$ to the Lamb shift in muonium. This correction is due  to insertion of radiative photons in the heavy line in the two-photon exchange diagrams. Our calculations are inspired by a new round of precise $1S-2S$ and $2S-2P$ experiments currently in progress.  
\end{abstract}

\maketitle


A new generation of precise $1S-2S$ and $2S-2P$ measurements in muonium and positronium is now either in progress or planned  \cite{Crivelli:2016fjw,Crivelli:2018vfe,Ohayon:2021dec,Mu-MASS:2021uou,Janka:2021xxr,Janka:2022pis,
Cortinovis:2023zqi,Zhang:2021cba,Uetake:2019,Iwai:2024tys,Blumer:2024fvc,Mu-MASS:2025zbg}. The Lamb shift and fine structure $2S-2P$ intervals in muonium have recently been measured with an unprecedented accuracy \cite{Mu-MASS:2021uou,Janka:2021xxr,Janka:2022pis,Mu-MASS:2025zbg}. Further reduction of the experimental uncertainty in measurements of these transition frequencies, as well as of the $1S-2S$ intervals is expected \cite{Cortinovis:2023zqi,Blumer:2024fvc}. This experimental progress requires improvement of the theory of spin-independent corrections to the muonium energy levels, see, e.g., \cite{Ohayon:2021dec,Janka:2021xxr,Cortinovis:2023zqi}. Triggered by the expected experimental results, some new spin-independent corrections to muonium energy levels were recently calculated \cite{Eides:2021wuv,Eides:2023ltp,Adkins:2022coe,Korobov:2024tlk}. 

\begin{figure}[h!]
\begin{center}
\includegraphics[width=10cm,height=1.8cm
]{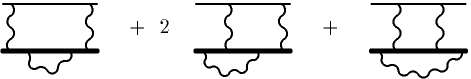}
\end{center}
\caption{Muon-line radiative-recoil corrections of order $Z^2\alpha(Z\alpha)^5(m/M)^2m$.
}
\label{muoneloop}
\end{figure}

\begin{figure}[h!]
\begin{center}
\includegraphics[width=10cm]{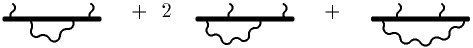}
\end{center}
\caption{One-loop diagrams for the muon factor.
}
\label{mufact}
\end{figure}

Below we calculate radiative-recoil contributions to the Lamb shift in muonium generated by the  gauge-invariant set of  diagrams in Fig.~\ref{muoneloop} (plus diagrams with crossed exchanged photons, which are not shown explicitly). Calculation of these diagrams is greatly facilitated by the use of the scattering approximation. Naively, one could expect that the diagrams
with a larger number of photon exchanges are not suppressed and could generate contributions of the same order as the diagrams in Fig.~\ref{muoneloop}. This is exactly what happens in the calculation of corrections of the order $Z^2\alpha(Z\alpha)^4(m/M)^2m$ (more on this later). However, this does not happen for corrections of the order $Z^2\alpha(Z\alpha)^5(m/M)^2m$. In this case, the contribution of the low ($\sim mZ\alpha$) loop momentum integration region in Fig.~\ref{muoneloop} is suppressed because the infrared behavior of any radiatively corrected Feynman diagram (or, more accurately, any gauge-invariant sum of Feynman diagrams) in Fig.~\ref{mufact} is softer than the behavior of the respective skeleton diagram\footnote{See some qualifications of this statement below.}. As a result, perturbation theory in $Z\alpha$ works in this case and addition of any extra exchanged photon always produces an extra power of $Z\alpha$. Hence, in calculations of the corrections of order $Z^2\alpha(Z\alpha)^5(m/M)^2m$ it is sufficient to consider  only the contributions of the diagrams in Fig.~\ref{muoneloop} in the scattering approximation (for more details, see, e.g., \cite{Eides:2000xc,Eides:2007exa}).

The hard spin-independent energy shift to the bound state energy level of two electromagnetically interacting leptons  generated by the diagrams with two-photon exchanges is described by the integral \cite{Eides:2000kj}

\beq  \label{general}
\Delta E=-\frac{(Z\alpha)^5}{\pi n^3}m_r^3
\int {\frac{d^4 k}{i\pi^2 k^4}} \frac{1}{4} Tr \left[(1 + \gamma_0 )L_{\mu \nu} \right]
\frac{1}{4} Tr \left[(1 + \gamma_0 )H_{\mu \nu} \right]\delta_{l0},
\eeq

\noindent
where $m$ and $M$ are the masses of the electron and muon, respectively, $L_{\mu \nu}$ and $H_{\mu \nu}$ are the light and heavy fermion factors, respectively, $m_r=mM/(m+M)$ is the reduced mass, $Z=1$ is the charge of the heavy fermion in terms of the positron charge, and $n$ and $l$ are the principal quantum number and the orbital momentum, respectively. The  expression in \eq{general} is exact in the mass ratio, and is valid also in the case of $m=M$ (positronium). 

The electron-line factor in \eq{general} has the form

\beq
L_{\mu \nu}=\gamma_{\mu} \frac{\slashed{p} + \slashed{k} + m}
{k^2 + 2mk_0 + i0} \gamma_{\nu}+
\gamma_{\nu}  \frac{\slashed{p} - \slashed{k} + m}{k^2 - 2mk_0 +
i0}\gamma_{\mu},
\eeq

\noindent
where $p=(m,{\bf 0})$ is the momentum of the particle with mass $m$.

\noindent
Calculating the electron trace in \eq{general} we obtain  

\beq \label{L0simple}
\frac{1}{4} Tr \left[(1 + \gamma_0) L_{\mu \nu}\right] 
=\frac{4m k_0^2}{k^4-4m^2k_0^2}\left[k^2g_{\mu 0} g_{\nu 0}\frac{1}{k_0^2}
- (g_{\mu 0} k_{\nu} + g_{\nu 0} k_{\mu})\frac{1}{k_0}+ g_{\mu \nu}\right].
\eeq

The expression for the muon factor is more complicated because it includes insertions of the radiative photons.  Both traces in \eq{general} have the inverse dimension of mass. Since the muon factor depends only on the heavy mass $M$, it is convenient to rescale the integration momentum $k\to Mk$. After rescaling, calculating the trace  and extracting the factor $Z^2\alpha/\pi$, which is due to radiative photons, the muon factor turns into 

\beq
\frac{1}{4} Tr \left[(1 + \gamma_0)H_{\mu \nu}  \right]\to\frac{1}{M}\frac{Z^2\alpha}{\pi}{\cal H}_{\mu\nu}(k),
\eeq

\noindent
where ${\cal H}_{\mu\nu}(k)$ is a function of dimensionless momentum k, and does not contain any mass parameters. This radiatively corrected muon factor is a sum of three terms 

\beq
{\cal H}_{\mu\nu}= {\cal H}_{\mu \nu}^{\Sigma} + 2{\cal H}_{\mu \nu}^{\Lambda}
+ {\cal H}_{\mu \nu}^{\Xi},
\eeq

\noindent
arising from the one-loop self-energy, vertex and spanning photon insertions in the muon line and 
corresponding to the diagrams in Fig.~\ref{mufact}.

The electron factor in \eq{L0simple} in terms of dimensionless momentum $k$ has the form 

\beq
\begin{split}
\frac{1}{4} Tr \left[(1 + \gamma_0) L_{\mu \nu} \right]&\to \frac{1}{M}\frac{4\mu k_0^2}{k^4-4\mu^2k_0^2}\left[k^2g_{\mu 0} g_{\nu 0}\frac{1}{k_0^2}
- (g_{\mu 0} k_{\nu} + g_{\nu 0} k_{\mu})\frac{1}{k_0}+ g_{\mu \nu}\right]\\
&\equiv \frac{1}{M}\frac{4\mu k_0^2}{k^4-4\mu^2k_0^2}{\cal L}_{\mu\nu},
\end{split}
\eeq

\noindent
where $\mu=m/M$ and ${\cal L}_{\mu\nu}$ is a dimensionless function, which does not contain any mass parameters.

The energy shift \eq{general} in terms of dimensionless  integration momentum $k$ is as follows

\beq  \label{shiftdiml}
\begin{split}
\Delta E&=-\frac{(Z^2\alpha)(Z\alpha)^5}{\pi^2 n^3}\frac{m_r^3}{M^2}\mu
\int {\frac{d^4 k}{i\pi^2 k^4}}\frac{4 k_0^2}{k^4-4\mu^2k_0^2}{\cal L}_{\mu\nu}(k){\cal H}_{\mu\nu}(k)\delta_{l0}\\
&=\frac{(Z^2\alpha)(Z\alpha)^5}{\pi^2 n^3}\frac{m_r^3}{M^2}\left(J_{\Sigma} + 2J_{\Lambda} + J_{\Xi }\right)\delta_{l0}.
\end{split}
\eeq

\noindent
We calculate the energy shift in \eq{shiftdiml} in the Yennie gauge for radiative photons. The integral in \eq{shiftdiml}  is linearly infrared divergent $\sim 1/k$ at small momentum $k$. This divergence is due to the one-loop muon factor  ${\cal H}_{\mu\nu}(k)$. All entries in the muon factor -- except the one-loop anomalous magnetic moment and one-loop slope of the electric form factor -- decrease fast enough with $k$ to make the integral in \eq{shiftdiml} infrared convergent. The slowly decreasing terms associated with the anomalous magnetic moment and the slope of the electric form factor produce the linearly infrared-divergent contributions and indicate existence of a contribution to the Lamb shift of the previous order in $Z\alpha$. 

The linear divergence is an artefact of the scattering approximation. In a full calculation it would be cutoff by the virtuality of the external lines at $1/k\sim 1/\gamma=M/(m_rZ\alpha)$ and/or $\mu/k\sim \mu/\gamma=m/(m_rZ\alpha)$. The terms $1/\gamma$ cancel  each other, while the terms $\mu/\gamma$ produce  contributions $Z^2\alpha(Z\alpha)^4(m/M)^2m$. These contributions of the previous order in $Z\alpha$ are well known \cite{Sapirstein:1990gn} and we simply omit linearly infrared-divergent terms of order $\mu/\gamma$. Individual diagrams in Fig.~\ref{muoneloop} generate also spurious logarithmic divergences $\sim\ln\gamma$. These divergences cannot be thrown away and   should cancel in the sum of the contributions of the diagrams in  Fig.~\ref{muoneloop}. The cancellation of logarithmic infrared divergences helps  keep control of the calculations.

Before proceeding further, let us mention that in a recent compilation \cite{Blumer:2024fvc} of the theory of muonium energy levels, the correction of order $Z^2\alpha(Z\alpha)^5(m/M)^2m$ is cited as 

\beq
\Delta E=\frac{Z^2\alpha(Z\alpha)^5}{n^3}\frac{m_r^3}{m^2}\left(\frac{m}{M}\right)^2
\left(\frac{139}{32}-2\ln2\right)\delta_{l0}. 
\eeq

\begin{figure}[h!]
\begin{center}
\includegraphics[width=10cm]{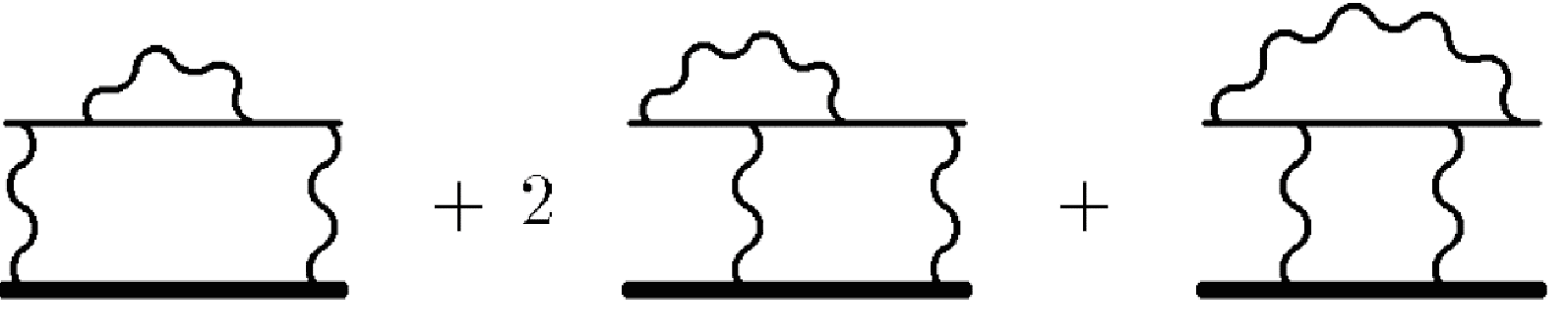}
\end{center}
\caption{Two-photon exchange diagrams with electron-line radiative insertions.
}
\label{elfactrad}
\end{figure}

\noindent
The last factor in the brackets hints at the origin of this  result. The classical nonrecoil correction of order $\alpha(Z\alpha)^5m$ is due to the two-photon exchange diagrams with radiative insertions in the electron line in Fig.~\ref{elfactrad}. In the external field approximation contribution of these diagrams to the energy shift is (see, e.g., review in \cite{Eides:2000xc,Eides:2007exa})

\beq \label{alp5nonr}
\Delta E=\frac{\alpha(Z\alpha)^5}{n^3}\frac{m_r^3}{m^2}\left(\frac{139}{32}-2\ln2\right)\delta_{l0}. 
\eeq

\noindent
Apparently, the authors of \cite{Blumer:2024fvc} noticed that formally the correction of order $Z^2\alpha(Z\alpha)^4(m/M)^2m$ can be obtained from the  leading contribution to the Lamb shift of order $\alpha(Z\alpha)^4m$ by the substitution $\alpha\to Z^2\alpha$, replacement in the argument of the logarithm $m\to M$ and  multiplication by $(m/M)^2$ (see, e.g., \cite{Sapirstein:1990gn}). While this recipe works for the leading contribution, it cannot be used for the corrections of the next order in $Z\alpha$   \cite{ohayon:2026}. 

All corrections of orders $\alpha(Z\alpha)^5m$ and $Z^2\alpha(Z\alpha)^5m$ originate from the high momentum integration regions in the diagrams in Fig.~\ref{elfactrad} and Fig.~\ref{muoneloop}. Notice that the factors $\alpha$ and $Z^2\alpha$ uniquely indicate  which set of the diagrams, in Fig.~\ref{elfactrad} or in Fig.~\ref{muoneloop} is responsible for this or that contribution. Besides the nonrecoil contribution in \eq{alp5nonr} the diagrams in Fig.~\ref{elfactrad} produce also linear (see, e.g., review in \cite{Eides:2000xc,Eides:2007exa})  and quadratic \cite{Blokland:2001fn} in the recoil factor $(m/M)$ contributions.

Our goal is to use \eq{shiftdiml} to obtain contribution of relative order $\mu^2$ to the energy shift. This is the leading in the recoil factor $(m/M)$ contribution of the diagrams in Fig.~\ref{muoneloop}. However, the factor before the dimensionless integral in \eq{shiftdiml} is already of  relative order $\mu^3$ and the contribution $\mu^2$ can arise only if the momentum integral in \eq{shiftdiml} (after omitting the linear divergences $\sim1/\gamma$) diverges as $1/\mu$ when  $\mu\to0$. Notice that the second order in $\mu^2$ contributions  to the muonium hyperfine splitting, which are due to the one-loop muon factor, have exactly the same origin \cite{Eides:1998vm}.

Further calculations are similar to the ones in \cite{Eides:2000kj,Eides:2021wuv,Eides:2023ltp,Eides:2022nda}. After deleting the terms of order $1/\gamma$ and preserving only the leading terms at $\mu\to0$ we obtain contribution of the diagram with the self-energy insertion in the muon line Fig.~\ref{muoneloop}

\beq \label{sigmaint}
J_\Sigma=-3 S_0 + 12S_2,
\eeq

\noindent
where $S_0$ and $S_2$ are the infrared divergent functions

\beq \label{asymps0}
\begin{split}
&S_0=
\mu\int_{\gamma^2}^\infty\frac{dk^2}{k^2}\frac{2}{\pi}\int_0^\pi d\theta\frac{\sin^2\theta}{k^2+4\mu^2\cos^2\theta}_{|\gamma<\mu\to0}\to 
\frac{2}{\gamma} + \frac{1}{\mu}\left(-\ln{\frac{\mu}{\gamma}} -1\right),\\
&S_2=
\mu\int_{\gamma^2}^\infty\frac{dk^2}{k^2}\frac{2}{\pi}\int_0^\pi d\theta\frac{\sin^2\theta\cos^2\theta}{k^2+4\mu^2\cos^2\theta}_{|\gamma<\mu\to0}\to 
\frac{1}{\mu}\left(\frac12\ln{\frac{\mu}{\gamma}} -\frac14\right).
\end{split}
\eeq

\noindent 
As mentioned above we will see that the term $\sim1/\gamma$ in \eq{asymps0} will cancel with similar terms from the other diagrams in Fig.~\ref{muoneloop}.

Calculation of the diagram with the vertex insertion in Fig.~\ref{muoneloop} is facilitated by the one-loop Yennie gauge expression for the renormalized  muon (electron) vertex  derived in \cite{Eides:2000kj,Eides:2001dw}. Soft infrared behaviour of the Yennie gauge vertex simplified calculations and we obtained  

\beq \label{spavetrtint}
J_\Lambda=\frac{\pi^2}{4}+ 3 S_0 - 12S_2+\frac{32}{3}\frac{\mu}{\gamma}\left(\ln\frac{1}{\gamma}-\frac{1}{3}\right).
\eeq

\noindent 
Notice, that in addition to the singular $1/\gamma$ and logarithmic in $\gamma$ terms in \eq{asymps0} the vertex insertion generates also singular terms of order $\mu/\gamma$.

For calculation of the diagram with the spanning photon  insertion in Fig.~\ref{muoneloop} we used a compact Yennie gauge expression for the last diagram  in Fig.~\ref{mufact} derived in \cite{Eides:2000kj}. After calculations we obtain   

\beq \label{spanonel}
J_\Xi=\frac{\pi^2}{2}- 3 S_0 + 12S_2-\frac{16}{3}\frac{\mu}{\gamma}.
\eeq

\noindent
This result contains the same singularities which are present in the vertex diagram.  

\noindent
The sum of the integrals in \eq{sigmaint}, \eq{spavetrtint} and \eq{spanonel} is

\beq \label{sumintall}
J_\Sigma+2J_\Lambda+J_\Xi=\pi^2+\frac{32}{3}\frac{\mu}{\gamma}\left(2\ln\frac{1}{\gamma}
-\frac{7}{6}\right). 
\eeq

\noindent
Notice that all linearly infrared divergent terms $\sim1/\gamma$,  as well as the  infrared logarithmically divergent contributions from the individual diagrams in Fig.~\ref{muoneloop} hidden in the functions $S_0$ and $S_2$, cancelled identically, as they should. The last linearly divergent term in \eq{sumintall} would generate the well known contribution $Z^2\alpha(Z\alpha)^4(m/M)^2m$  of the previous order \cite{Sapirstein:1990gn} and we omit it. All finite contributions proportional to $\pi^2$ in \eq{sumintall} originated from the vertex and spanning photon diagrams in Fig.~\ref{muoneloop}, while such contributions from the self-energy diagram cancelled each other. This is an accident cancellation which is due to the Yennie gauge used in the calculations above. 

Finally, we obtain the total radiative-recoil contribution to the Lamb shift of order $Z^2\alpha(Z\alpha)^5(m/M)^2m$  

\beq \label{mufactr}
\Delta E_\mu=\frac{(Z^2\alpha)(Z\alpha)^5}{\pi^2 n^3}\frac{m_r^3}{M^2}\left(J_\Sigma+2J_\Lambda+J_\Xi\right)\delta_{l0}
=\frac{(Z^2\alpha)(Z\alpha)^5}{n^3}\frac{m_r^3}{M^2}\delta_{l0}.
\eeq

The radiative-recoil correction of order $\alpha(Z\alpha)^5(m/M)^2m$ due to the radiative photon insertions in the electron line in Fig.~\ref{elfactrad} was calculated in \cite{Blokland:2001fn}

\beq \label{elfactr}
\Delta  E_e=\frac{\alpha(Z\alpha)^5}{n^3}\frac{m_r^3}{M^2}\left(-\frac{127}{32}+8\ln2\right)\delta_{l0}.
\eeq

\noindent
The factor $Z=1$ in muonium, so it makes sense to combine our new result in \eq{mufactr} and the result in \eq{elfactr} to obtain a total contribution to the energy shift of relative order $(m/M)^2$

\beq \label{sumelmu}
\Delta E_t=\frac{\alpha^6}{n^3}\frac{m_r^3}{M^2}\left(-\frac{95}{32}+8\ln2\right)\delta_{l0}.
\eeq

\noindent
Numerically it contributes 0.97 kHz to $1S-2S$ transition frequency in muonium, to be compared with the goals of MU-MASS \cite{Crivelli:2018vfe} and J-PARC \cite{Iwai:2024tys} experiments to reduce the experimental uncertainty below 10 kHz. We expect the theoretical results in \eq{mufactr} and \eq{sumelmu}, as well as other recent theoretical results \cite{Eides:2021wuv,Eides:2023ltp,Adkins:2022coe,Korobov:2024tlk}, to find applications in the discussion of the forthcoming experimental results on the muonium energy levels.

\acknowledgments

Work of M. I. Eides was supported by the NSF grant PHY-2510100.

\end{document}